\documentclass{article}

\usepackage{graphicx}

\begin{document}

\title{\textbf {Émilie du Châtelet and Euler: A Rare Convergence on the Hypotheses of Physics}}
\author{Dora Musielak}
\date{}

\maketitle

\begin{abstract}
Euler stressed the importance of hypotheses, which he thought were the only means of arriving at a certain knowledge of the physical causes, essential to establish the laws of physics. This thought was communicated to Émilie du Châtelet in response to hers when she debated the nature of forces, defending the Leibnizian concept of {\it vis viva}. After examining the Euler-Châtelet correspondence, I introduce a treatise discovered in 1844 where Euler provides the first analytical attempt to explain the difference between momentum and {\it vis viva}, and where he defined a new concept related to motion that can be considered the first idea of kinetic energy. These documents have received little attention. In this paper, I examine the topics Euler discussed in these manuscripts and place them in a context within the scientific and philosophical research of the seventeenth and eighteenth centuries that sought to establish the principles of nature, and that served as foundation for physics.

\end{abstract}



\section{Introduction}

The Marquise du Châtelet, like other thinkers of the Enlightenment, aspired to understand the exact sciences and gained a reputation for contributing to expound the scientific principles derived from philosophical thought. She studied the nascent science of mechanics through the philosophy of Newton, Descartes, and Leibniz, and in the process, she participated in the debate that raged in the eighteenth century between Cartesians and Newtonians, as she promoted the {\it vis viva} (living force) concept introduced by Leibniz to differentiate the force of a body which has only the tendency to motion, as opposed to the force when the body is actually moving.

In 1963, a volume of correspondence written by Euler to several scientists was published.\footnote{Smirnov 1963.} There, among letters to Clairaut there is a copy of an undated draft addressed to Châtelet that provides the historical and scientific background for the research in this paper. See Appendix A for my annoted copies.
\subsection{Émilie du Châtelet: Newtonian Physicist}

The Marquise du Châtelet (also written Chastelet, or Chastellet\footnote{Châtelet is a modern form derived from the original French name Chastelet. The spelling was fixed in the nineteenth century. Before that time, the name of the Marquise was spelled Chastelet or Chastellet. The circumflex on the “a” stands for an “s” after this vowel. This also occurs in “Hôpital” from the old French “Hospital.”}) never met Isaac Newton, as he died before she learned of his theories. However, she was perhaps the first woman in history who mastered Newtonian physics. At twenty-eight, she began to study with Pierre Louis M. de Maupertuis, Alexis Clairaut, and later with Johann II Bernoulli (son of Johann I). In 1738, she engaged Samuel König to give her private lessons to learn the metaphysics of Gottfried W. Leibniz and Christian Wolff.\footnote{For details on Châtelet’s personal life and her studies, Cf. Musielak 2014; Hammel 1910.}

To contribute to the scientific enlightening of her contemporaries, Châtelet made the first translation of Newton’s {\it Principia} into French.\footnote{Châtelet 1759.} Her intention was to make Newton’s theories accessible and better understood, and in doing so, she contributed to the widespread campaign to promote the Newtonian philosophy of natural science. 

Émilie du Châtelet entered the world of science at a crucial time when natural philosophers sought to establish new rules that would reduce rational mechanics to set of irrefutable axioms or a single principle from which all others could be derived. Savants of the eighteenth century searched for deeper understanding of physics, asking basic questions such as, {\it What is the correct measure of force of a body in motion?}, {\it What motion quantity is conserved?} Châtelet expounded the most important theories of her time, aiming to answer those questions.

\subsection{Leonhard Euler: Mathematicorum Principi and Most Eminent Authority in Mechanics}

In 1740, Leonhard Euler was at the forefront of the effort to more precisely establish mechanics using the full power of analysis. Educated on Newtonian mechanics and the Leibnizian calculus, he was an expert on Newton’s and Descartes’s natural philosophy. Above all, Euler was the main developer of the analytical calculus tools used in all branches of physics. While at the Russian Academy of Sciences in Petersburg, Euler published {\it Mechanica sive motus scientia analytice exposita},\footnote{Euler 1736.} his first major treatise where he used calculus for the first time to express Newtonian mechanics in analytic form, beginning the unified structure for analytical mechanics.

A gifted mathematician, Euler believed that knowledge is founded (in part) on the basis of precise quantitative laws. He insisted that mathematics preparation was crucial to pursue fundamental studies of physical phenomena. Euler accepted and extended Newton’s dynamics, while also incorporating elements from the Cartesian and Leibnizian sciences and his own natural philosophy.

Much has been written about Euler’s monumental contribution to all the exact sciences, and thus I will not repeat but just the essential facts to help me illuminate the arguments herein. I will focus on the ideas he set forth in the short correspondence he had with the Marquise du Châtelet.

\section{Descartes, Newton, and Leibniz: Search for Physical Principles}

In the 1730s, the scientific climate in Europe brewed with ideas from two opposite groups of thinkers: Cartesians and Newtonians. René Descartes had provided the first distinctly modern formulation of laws of nature and a conservation principle of motion, and he also constructed what would become the most popular theory of planetary motion of his time based on his model of vortices.

The Cartesian model of the universe was full of matter and the concept of vortices served to explain the orbits of the planets about the Sun with the heavier objects spinning out towards the outside of the vortex and the lighter objects remaining closer to the center. Then with his {\it Principia Mathematica}, Isaac Newton stated the laws of motion and a new principle of gravitation to explain how our Solar System works that contradicted the Cartesian hypothesis of vortices. However, Gottfried Leibniz criticized Newton for not explaining how gravity acts across space, while French scientists argued that the force of gravity was merely a supernatural idea. The ideas put forward by Descartes, Newton, and Leibniz (Fig. 1) were vigorously debated.

\begin{figure}
\begin{center}
\includegraphics{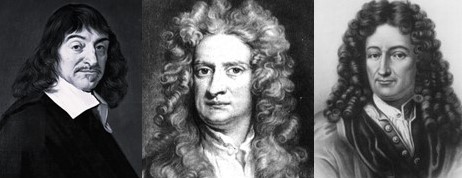}
\end{center}
\caption{René Descartes (1596-1650), Isaac Newton (1643-1727), and Gottfried Leibniz (1646-1716)}
\label{figure}
\end{figure}

\subsection{The Vis Viva Controversy}

In the seventeenth and eighteenth centuries, scientific and philosophical research sought to establish the laws of nature. In 1644, Descartes published {\it Principles of Philosophy}, in which he attempted to deduce scientific knowledge from metaphysics. The {\it Principles} consists of four parts: {\it The Principles of Human Knowledge}, {\it The Principles of Material Things}, {\it Of the Visible World}, and {\it The Earth}; Descartes intended to put the whole universe on a mathematical foundation, reducing the study to one of mechanics. The Cartesian universe was filled with matter which, due to some initial motion, had settled down into a system of vortices which carry the Sun, the stars, the planets and comets in their paths, a model known as “theory of vortices” ({\it théorie des tourbillons}). The {\it Principles} was the Cartesian celestial mechanics that influenced the best scientific minds of the seventeenth and eighteenth century, including Leibniz, Newton, the Bernoullis, Euler, and others.

Descartes presented three laws of nature, defined the quantity of motion, and provided a full explanation of their metaphysical background and physical consequences. In his view, God, the general cause of all motion in the universe, preserves the same quantity of motion and rest put into the world at the time of creation. By stating that “we must reckon the quantity of motion in two pieces of matter as equal if one moves twice as fast as the other, and this in turn is twice as big as the first,”\footnote{Descartes 1644.} Descartes implied that the measure of the quantity of motion is the product of mass and velocity ($mv$).

Leibniz discovered what he considered to be a serious flaw in Descartes’s conservation law, namely, that it violates the principle of the equality of cause and effect (i.e. conservation law). In 1686, Leibniz published in Acta Eruditorum {it A Brief Demonstration of a Notable Error of Descartes’s and Others Concerning a Natural Law},\footnote{Leibniz 1686.} a note where he rejected the Cartesian equivalence between motive force, which Leibniz agreed is conserved in nature, and quantity of motion, which he argued is not.

In 1695, Leibniz published a memoir on his new science of dynamics: {\it Specimen dynamicum}. In this work, Leibniz introduced {\it vis viva} to differentiate between living and dead forces. As examples of dead forces, he gave centrifugal force and gravitational or centripetal forces; Leibniz also included the forces involved in static equilibrium that, when unbalanced, initiate motion.

Leibniz asserted that the quantity which remains absolute and indestructible in the universe is not the Cartesian quantity of motion but {\it vis viva} or living force ($mv^2$). Leibniz ideas ignited a fierce dispute among natural philosophers, forcing a profound and much contested debate that become known as the “vis viva controversy.”\footnote{Cf. Terral 2004; Smith 2006.} Leibniz’s concept of {\it vis viva} clearly opposed the ideas of Descartes regarding motion and also appeared to oppose Newton’s theory of conservation of momentum, which Descartes had advocated.

The conservation of motion conceived by Descartes was accepted as it was believed that all space is filled with matter. Leibniz agreed on the conservation of directional motion, but argued that because it is directional, unlike $mv^2$, it is not conserved in a collision.

Referring to the metaphysical principle (that the effect must equal the cause), Leibniz calculated “the force through the effect produced in using itself up” and concluded that the force transferred from one equal body to another varies as the square of the velocity. For Leibniz the metaphysical principle established the priority of the conservation of living forces in changes of motion.

\subsection{Prize Competition Attracts Euler and Châtelet}

The French Academy of Sciences was instrumental in fostering research and promoting advances in all areas of the exact sciences by offering prize competitions where important questions were posed to scholars. It is not surprising that since the inception of the prize in 1720, the topics concerned the principles of motion addressing the general system of the world, and physical astronomy. The first question posed was “What is the principle and nature of motion, and what is the cause of the communication of motions?” followed by a demonstration of the laws of impact of bodies in 1724, and the laws of impact of bodies with a perfect or imperfect spring in 1726. By 1728, scholars were still unsure on the causes of gravity or what causes the elliptical shape of planetary orbits.\footnote{Musielak 2022.} The leading mathematicians answering those questions included Johann I and Daniel Bernoulli and Colin Maclaurin. After seeking answers on light propagation, the Academy prize for 1738 concentrated on the nature and propagation of heat.

Driven by her scientific ambition, Madame du Châtelet wrote a 139-page {\it Dissertation sur la nature et la propagation du feu} [“Dissertation on the nature of fire”] and submitted it for the Academy’s competition. As identifier of her anonymous entry, she chose a most appropriate epitaph: {\it Ignea convexi vis \& sine pondere coeli Emicuit, summaque locum sibi fecit in arce}.\footnote{from Roman poet Ovid’s Methamorphoses, Book 1, lines 26-27, which can be translated as “the fiery ({\it ignea}) and weightless ({\it et sine pondere}) force ({\it vis}) of the arching sky ({\it convexi caeli}) emited ({\it emicuit}) and made ({\it …que fecit}) for itself ({\it sibi}) a place ({\it locum}) in the highest citadel ({\it summa in arce}).}
 
In this essay Châtelet attempted to make a synthesis of all knowledge on the nature of fire. Of course, she didn’t have new experimental facts nor did she offer a valid scientific theory for propagation of heat. Her introduction was rather weak, starting her {\it Dissertation} with a simple statement: {\it Le feu se manifeste à nous par des phénomènes si différents, qu’il est aussi difficile de le définir par ses effets, qu’il paraît impossible de connaître entièrement sa nature}. [“Fire is difficult to define and it is impossible to fully understand its nature.”] She concluded the first part by saying that light and heat are two different and independent effects. To discuss the propagation of fire, she also addressed the nature of the Sun, saying, {\it Le Soleil ne peut être un globe de feu}, [“The Sun cannot be a globe of fire”], quoting Newton extensively to describe the colors of sunlight. Despite her effort, it did not win.

The 1738 Academy’s prize was awarded to 31-year-old Leonhard Euler,\footnote{Euler 1738.} and to two other lesser-known scholars.\footnote{Comte de Créquy, and Pére Louis-Antoine Lozeran de Fiesc de la Campagnie de Jesus.} With his {\it Dissertatio de igne} (the only entry written in Latin), Euler proposed that fire is the result of the bursting of tiny glassy balls of highly compressed air in the pores of bodies, so that “heat consists in a certain motion of the smallest particles of a body.” Euler believed that all the phenomena associated with heat and fire could be deduced from the laws of mechanics without presupposing any “occult qualities.” Euler stated, “light is the elastic vibration of the {\it ether} that is initiated by the explosions of little balls; hence, light is propagated by the same laws as sound.”\footnote{Cf. C. A. Truesdell’s Introduction to Opera Omnia Series.}

Du Châtelet was rather annoyed at the results of the competition. Was she expecting a favorable review since Maupertuis was in the Academy’s review committee (commissaire pour les prix)? She complained. On 28 May 1738, she wrote to Maupertuis:  “{\it Nous sommes au désespoir en voyant le jugement de l’Académie (des Sciences); il est dur que le prix ait été partagé, et que M. de Voltaire n’ait pas eu part au gâteau. Ce M. Fuller [sic] qui est nommé, est un Léibnitien, et, par conséquent, un Cartésien : il est fâcheux que l’esprit de parti ait autant de crédit en France.}"\footnote{Châtelet 1818. p. 38.}

She was referring to Leonhard Euler. Her note seems to refer to Euler with disdain, believing that Voltaire had more merit than the young Euler! Clearly, Madame du Châtelet should have known that Euler was a true scientist, the greatest mathematician of her time.\footnote{In the 1730’s, Euler had established his international fame by solving important, long-standing problems in almost all fields of contemporary mathematics. Euler’s 1736 {\it Mechanica} was used extensively in the 1741 edition of Newton’s {\it Principia mathematica} by Le Seur and Jacquier.}

Thanks to the favorable influence of mutual friends at the Academy of Sciences, Châtelet’s essay was published together with the winning memoires. The inclusion was justified by stating that the “authors’ names were likely to arouse the interest and curiosity of the public.” \footnote{Pièces qui ont remporté le prix de l’académie royal des sciences de Paris, 1738. pp. 85-86.}

And this is how Euler became acquainted with the gifted Marquise du Châtelet. On 19 February 1740, Euler wrote: “It is an honor, Madam, to see myself in a competition with a person who makes one of the rarest ornaments of her sex, by the brilliance with which you illuminate the most exalted sciences, by bringing to them the sublimity of your genius.”\footnote{See Letter 1 in Appendix A, from Smirnov 1963, p. 281.}

\section{Institutions de Physique (1740): Synthesis of Physical Theories and Philosophical Doctrines}

Having her essay on the nature of fire published by the Academy of Sciences gave Émilie a sense of accomplishment, and it helped to invigorate her scientific studies. She sought and obtained the approval to publish a treatise intended to present the philosophical principles of Newton. Her work was ready as early as 18 September 1738, but she delayed its publication to include the metaphysics of Leibniz.

Finally, her {\it Institutions de physique} (Physical Institutions)\footnote{Châtelet 1738-1740.} was published anonymously in 1740, although Châtelet made sure everyone knew she was the author. In this treatise, she addressed the principles of physics and mechanics based on Newton’s work, and expounded the philosophical work of Leibniz and the doctrines of Christian von Wolff.

At that time, the pursuit of natural laws was divided into competing Newtonian, Cartesian, and Leibnizian formulations, each incorporating metaphysics, natural philosophy (physics), and mathematics.\footnote{Metaphysics was used to reinforce particular theories or doctrines.} Hence, Châtelet’s {\it Institutions} aimed at presenting a united front, a lofty goal that could not be achieved because many concepts in those distinct formulations were incompletely defined or simply not well understood.

The {\it Institutions} began with topics of metaphysics (e.g. principles of human knowledge, existence of God) before introducing fundamental concepts of physics (e.g. space, time, elements and divisibility of matter), and then it addressed motion, gravity, and Newtonian and Cartesian physics, including the research by Galileo, Huygens, Johann I Bernoulli, Jean-Jacques Dortous de Mairan, and others.

Châtelet claimed that this treatise was intended for the education of her young son, Louis-Marie-Florent, then thirteen. However, only scholars could appreciate her book. As Clairaut remarked, “it would be difficult for beginners and especially for lay people, who unfortunately would be the ones who would judge her the most.”\footnote{{\it Mais je crains que cela ne soit difficile aux commençants et surtout aux gens du monde. Malheureusement ce sont ceux qui vous jugeront le plus et qui s'en prendront à vous et non à eux comme ils le devraient de ce que vous vous distinguez.} Letter from Clairaut to Chatelet, 4 Jan 1741.} Clairaut was right.

Émilie du Châtelet covered the topics that were the subject of controversy at the time: the nature of forces. She gave credit to Leibniz for discovering the vital forces, i.e. to have “guessed one of the secrets of the creator.”\footnote{Letter from Châtelet to Maupertuis dated February 1738.} Châtelet presented the Physical Institutions as a “small sketch” of the metaphysics of Leibniz and his disciple Wolff, who advocated for Leibniz’s theory of monads as expounded in {\it The Monadology}, the metaphysics of simple substances, or monads.

Proud of her work, Châtelet sent copies of the {\it Institutions} to educated acquaintances. She wanted to be recognized for her cleverness in the sciences as in her understanding of natural philosophy. Driven by such desire, she also sent her treatise to Frederick II, King of Prussia, who was a friend of Voltaire. Because Frederick did not like Émilie, he did not care for the gift. The king wrote to Voltaire to praise his work on Newtonian philosophy, while at the same time downplaying hers: “When I read {\it les Institutions de physique} of the marquise, I do not know if I have been deceived or if I am deceiving myself.”\footnote{Davidson 2012, p. 168.} Johann Bernoulli was also critic of her work. In a letter of 12 April 1739, Bernoulli scolded Maupertuis for “not instructing Madame du Châtelet more precisely in the theory of {\it les forces vives}.”\footnote{{\it Je m’étonne, Monsieur, que depuis si longtemps que vous connoissez cette Dame philosophe, vous ne lui ayez pas donné de meilleures instructions sur cette importante matière.} Brown 1975, p. 90.}

\section{Châtelet-Mairan Debate Over Forces (1741)}

With {\it Institutions}, Châtelet stirred the debate related to {\it vis viva} or “living force”. This concept served as an elementary and limited formulation of the conservation principle of motion. It was the first description of what we now call kinetic energy but, as defined then, it was misunderstood and {\it vis viva} caused a debate regarding Newtonian mechanics and the concept of force.

\subsection{Opposing Views}

Châtelet was aware of the disagreement about Leibniz’s concept of {\it vis viva} and the ideas of Descartes and Newton regarding motion. During the period 1676 – 1689, Leibniz concluded that in many mechanical systems (of several masses, $m_i$ each with velocity $v_i$) the quantity $\sum m_i v_i^2$ is conserved. He called this quantity the {\it vis viva} or living force of the system. This principle can represent the conservation of kinetic energy in elastic collisions, which is independent of the conservation of momentum.

But scholars of the eighteenth century did not distinguish momentum and energy of motion. Many were influenced by Newton’s prestige in England, and Descartes’s authority in France, both of whom had promoted the conservation of momentum as a guiding principle. In other words, they believed the momentum $\sum m_i v_i$ was the conserved quantity of motion.

Thus, the issues that Madame du Châtelet grappled with required establishing the quantities that are universally conserved: Leibniz’s {\it vis viva}, or Descartes’s momentum. The problem was controversial because it involved several other issues. One was semantic: the interpretation of “force” was not clear. The other was the metaphysical issue raised by Leibniz. And of course, the empirical issue of the apparent nonconservation of {\it vis viva} in the collision of soft bodies was even less clear.

Châtelet embraced {\it vis viva}: [in the {\it Institutions}] “I wanted to give an idea of the metaphysics of Leibniz that I confess to being the only one that has satisfied me, even though I still have doubts.”\footnote{Letter from Châtelet to Maupertuis dated 29 September 1738.}

\subsection{On Forces}

Chapter XX of the {\it Physical Institutions} was devoted to dead forces ({\it forces mortes}) and the equilibrium of powers ({\it equilibre des puissances}). Châtelet distinguished two types of forces: dead force or virtual force ({\it Force morte ou Force virtuelle}), and living force ({\it Force vive}). Dead force, she said, consists of a simple tendency to movement: such is that of a spring ready to relax; and the living force is that which a body has when it is in an actual motion.\footnote{Châtelet 1740, §. 519, p. 399.} She added that dead forces are also called pressing forces ({\it Forces pressantes}), because they press the bodies that resist them, and they are an effort to disturb them from their position.\footnote{Ibid. §. 520, p. 399.} Later on Châtelet stated that the measure of dead force is the product of the mass times the initial velocity.\footnote{Ibid. §. 561, p. 415.}

Chapter XXI, {\it de la Force des Corps}, begins by stating that “A body cannot pass suddenly from motion to rest, nor from rest to motion.” Then it goes on to affirm that “All mathematicians agree on this principle, they always measure the ratio of efforts or dead forces by the products of the masses multiplied by the initial velocities, and no one has ever thought of calling this truth into doubt; but it is not the same with {\it vis viva}, that is to say, with the force which resides in a body which is in an actual motion, and which has a finite speed, that is to say, say, a speed infinitely greater than this initial speed of which I have just spoken.”\footnote{Ibid. §. 565, p. 419.} Châtelet credits Leibniz for discovering “the true measure of {\it vis viva} and for distinguishing the two forces.”

\subsection{Arguments on Concept of Force}

In Chapter XXI of her {\it Institutions}, Châtelet questioned the theory of forces promoted by Jean-Jacques d’Ortous de Mairan.\footnote{De Mairan’s work includes contributions to theory of heat, observations of meteorological phenomena, and on the orbital motion and rotation of the Moon. He noted a small nebulosity around a star closely north of the Orion Nebula.} Mairan, at that time the Secretary of the French Academy of Sciences, was a friend who had helped her in publishing her work. In assessing de Mairan’s essay on estimation and measurement of the driving forces of the bodies (published in 1728), Châtelet criticized it for advocating that the strength of the body corresponds to the simple product of mass and velocity (Cartesian idea).

Châtelet stated that, in determining the force of a body, the correct definition was that given by [Johann I] Bernoulli and Leibniz. In other words, the strength of the body should be the product of its mass by the square of its velocity. As soon as du Châtelet’s book appeared in print, her critical assessment of de Mairan’s theory of forces caused a heated dispute with him. De Mairan published a long essay regarding the question of {\it les forces vives}, to respond to her harsh critique.\footnote{M. de Mairan, Secrétaire Perpétuel de l’Académie Royale des Sciences, à Madame la Marquise du Chastellet sur la question des forces vives, en réponse aux objections qu’elle lui a fait sur ce sujet dans les Institutions de Physique.} She fired back.\footnote{Réponse de Mme *** à la lettre de M. de Mairan sur la question des forces vives, J. de Savants, 1741. Her response is dated 18 February 1741.}

Both contenders adhered to what they believed to be the true physical concept. However, neither one could move beyond what had already been established by their predecessors. Newton for example had failed to define force. In {\it Principia} he states: “… I offer this work as the mathematical principles of philosophy, for the whole burden of philosophy seems to consist in this—from the phenomena of motions to investigate the forces of nature, and then from these forces to demonstrate the other phenomena.”\footnote{Newton’s Preface to the first edition of Principia, p. xvii. A slightly different wording for this passage is given by Cohen and Whitman in their 1999 translation, p. 382.} Newton’s second law asserts that “the change of motion is proportional to the motive force impressed, and it takes place along the right line in which that force is impressed.” It does not tell us what is the measure of “motion” or how to determine forces.

Leibniz believed in the conservation of {\it forces vives} ($mv^2$). His {\it vis viva} represents a measure of a body’s ability to bring about effects by virtue of its motion. It is an active force which allows a moving body to, say, raise itself up to a given height or impart a motion to a slower body. The {\it vis viva} [force] that a body transfers to another body through impact must be equal to the {\it vis viva} it loses during that impact.

For Châtelet, there were two [body] forces: dead force and living force. As she defined, dead force was the simple tendency to movement, such is that of a spring ready to relax and thus she also called it virtual force; living force she took as that which a body has when it is in actual movement, that is Leibniz's {\it vis viva}, $mv^2$.

However, for de Mairan, force and motion were nothing but magnitudes susceptible of increase or diminution … measured force lost during unit of time … when a body comes to rest, all motive force will be used up and the amount will equal the initial force of the body. For retarded motion, he believed that the measure of force was the amount of momentum, $mv$, lost in each interval.

Today, we use two formulae: $m_iv_i$ for momentum and $\frac{1}{2} mv^2$ for kinetic energy.\footnote{Smith 2006.} Did Châtelet understand the difference between mass ($m$) and weight ($mg$), or between power ($Fv$) and momentum ($mv$), or between momentum ($mv$) and {\it forces vives} ($mv^2$)? Did she grasp that {\it vis viva} and momentum were equally valid concepts to describe motion? It would be unfair to expect that she did, considering the state of knowledge that she had at her disposal.

\subsection{Institutions Physiques, 1742}

Châtelet was fully convinced that her views on forces were correct. In the spring of 1742, she published a second edition of her treatise under a slightly altered title and shows the name of its author: {\it Institutions physiques de Madame la marquise du Châstellet adressés à Mr. son fils}.\footnote{Châtelet 1742.} This time she also included her portrait, (Fig. 2) with a proud epitaph below it: “This is how the Truth better establishes its power, taking on the features of beauty and the graces of Eloquence.”

Clairaut, her friend and teacher, served as referee of her original {\it Institutions}. In the spring of 1741, he sent her a number of comments to help her improve the text and sugestions to clarify some concepts. As noted by the printer, although she had made many changes to her work for the second edition, she made none to the chapter on forces, Chapter XXI, (only a few words were added to §. 582 to clarify it), so that the “Reader will find it here as it was when the public dispute she had with Mr. De Mairan, about the force vives, began. Attached to this Edition is the letter of Mr. De Mairan to the Author of the Institutions, and her Reply to him; these are so far the only two pieces that this dispute has produced.”\footnote{Châtelet 1742, pp. 435-436.}

\begin{figure}
\begin{center}
\includegraphics{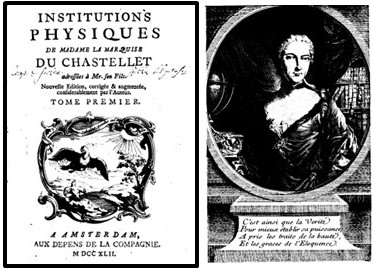}
\end{center}
\caption{Gabrielle Émilie Le Tonnelier de Breteuil Marquise du Châtelet (1706-1749). Frontispiece of {\it Institutions Physiques}, MDCCXLII (1742).}
\label{figure}
\end{figure}

\subsection{Seeking Euler}

During her dispute with de Mairan, Châtelet sought Euler’s endorsement when neither Maupertuis nor Clairaut, her closest friends and mentors, came to her defense. Before that, Euler had written her a short congratulatory note dated 19 February 1740 (Julian calendar).\footnote{Russia still used the Julian calendar while the rest of Europe used the Gregorian calendar. When it was 19 February in St. Petersburg, it was already March 1 in Paris.} In this letter,\footnote{See Letter 1 in Appendix A.} a very polite Euler, who was still in St. Petersburg, wrote to thank Châtelet, noting that he had not received the {\it Pièce} that she had sent him through Maupertuis. The {\it pièce} referred to her 1740 {\it Institutions de Physique}. In 1741, Châtelet asked Maupertuis to forward a copy to Euler, hoping that he would help her appease her critics.

In any case, in 1740 Euler could not comment on her {\it Institutions} since he had not read it. However, he praised du Châtelet anyway, as he knew her work on the nature of fire published two years earlier (together with his prize-winning dissertation). Euler added:

{\it ... Il m’est bien glorieux, Madame, de me voir en lice avec une personne qui fait un des plus rares ornements de son Sexe, par le Lustre que Vous avez bien voulu répandre sur les sciences les plus relevées, en y portant la Sublimité de Votre génie. Cela seul, je le sens bien, est capable de me donner un Relief que je n’oserais espérer de mes faibles lumières ...} [“It is a great honor for me, Madame, to find myself in a competition with a person who makes one of the rarest adornments of her sex, by the brilliance with which You illuminate the most exalted sciences, by bringing to them the Sublimity of Your genius. This alone, I feel it is capable of giving me the depth that I would not dare to hope for from my modest knowledge.”]{\footnote{Smirnov 1963, pp. 280-281. See Letter 1 in Appendix A.}}

In the summer of 1741, the public debate between Châtelet and de Mairan was raging. In June, after learning that Euler was on his way to Berlin, Châtelet wrote to Maupertuis that she wanted Euler to read her {\it Institutions}, including her letter to de Mairan. On August 8, Émilie wrote again (she was in Brussels), after learning from the newspapers that Euler had arrived to Berlin.

In an undated and unfinished letter to Châtelet,\footnote{Smirnov 1963. pp. 275-279.} Euler addressed her {\it Institutions}, concentrating in particular in her chapter on hypotheses. He was pleased to learn that she fought strongly and firmly with those philosophers who wanted to banish completely the hypotheses in physics that were, in his opinion, the only way to achieve a certain knowledge of the physical causes. Châtelet asserted in her book that physical laws had to be understood through hypotheses.

On 8 May 1742, Euler wrote to Christian Goldbach and mentioned receiving copy of the new edition of Châtelet's {\it Institutions} and her portrait.\footnote{Die Madame La Marquise du Chatellet hat mir ein Exemplar von der neuen Edition der Institutions Physiques nebst ihrem Portrait zugeschickt. See Euler (2015), p. 187.} He made no comment about its merits. In response, Goldbach mentioned that the Chevalier de Mouhy would be writing a review. However, no published review by Mouhy\footnote{Charles de Fieux Chevalier de Mouhy (1701-1784) was a French journalist and homme de lettres, author of novels and essays on cultural history; he was a supporter of Voltaire and thus of Châtelet.} of the {\it Institutions} has been found to date.

\section{Topics Euler Addressed in (Undated) Letter}

Euler begins the undated letter by praising the {\it Physical Institutions}, impressed by the clarity with which she treated this science [physics], and the ease with which she had explained the most difficult ideas about motion. But above all, Euler was most pleased with her Chapter on hypotheses.

Then Euler proceeds to discuss {\it matière subtile}, a topic that is not part of Chatelet’s {\it Institutions}, referring in particular to Musschenbroek,\footnote{Pieter van Musschenbroek (1692 – 1761) was a Dutch scientist, discoverer of the principle of the Leiden jar. He was a professor in Duisburg, Utrecht, and Leiden, where he taught mathematics, philosophy, medicine, and astrology. His Elementa Physica (1726) played an important part in the transmission of Newton’s ideas in physics to Europe. Other important works: Dissertationes physicae experimentalis et geometricae de magnete (1729); Institutiones physicae (1734).} a contemporary physicist that she never mentioned in her treatise either. Thus, this part of Euler’s letter is puzzling, and I argue that he may have used this passage to critique her support for Leibniz and the monadology.

After lamenting that the main topic has been already too long the subject of a bitter dispute, “whereby mathematics has lost much of its reputation,” Euler finally responds to Chatelet’s request to address the concept of forces. 

In the following sections, I provide additional background to better understand Euler’s message, focusing on the main ideas: Hypotheses of physics, subtle matter, the forces of bodies, and the first principle of mechanics.

\subsection{Hypotheses of physics}

For Euler, “the hypotheses of physics are the only means of arriving at a certain knowledge of the physical causes.”\footnote{Euler to Châtelet; see Letter 2 in Appendix A.} Châtelet agreed. She wrote that the true causes of the natural effects and the phenomena that we observe are often so far removed from the principles on which we can rely, and from the experiments that we can make, that we are obliged to content ourselves with probable reasons to explain them: probabilities are not only because they are often of great use in practice, but also because they clear the way and lead us to the truth.\footnote{Châtelet (1742), §. 53.}

Euler dully praised Châtelet’s treatment of hypotheses: “But above all, the Chapter on Hypotheses gave me the greatest pleasure, seeing, that You are fighting, Madam, so strongly and so solidly some {\it English Philosophers} who want to banish altogether {\it the hypotheses of Physics which are nevertheless in my opinion the only way to achieve a certain knowledge of the physical causes}.”\footnote{Smirnov 1963. See undated Letter 2 in Appendix A.}

Who were the “English Philosophers” who were against hypotheses? Was Euler referring to Newton himself?  In the 1713 (second) edition of his {\it Principia}, Newton added a General Scholium which served him to discuss God and hypotheses, and where the now famous sentence {\it Hypotheses non fingo} first appeared. The Scholium was a form of conclusion to Book 3 where Newton explained the phenomena of the heavens and of the sea by the force of gravity, admitting that he “had not yet assigned a cause to gravity.”  {\it Hypotheses non fingo} is unquestionably the most legendary of Newton’s dicta, and for decades scholars have debated its meaning, often disagreeing on the translation and its actual meaning. 

A 1999 translation of the {\it Principia} presents Newton’s statement as follows:\footnote{Newton 1726. See Book 3, General Scholium, p. 943.}
\smallskip

“I have not as yet been able to discover the reason for these properties of gravity from phenomena, and {\it I do not feign hypotheses}.\footnote{The word “fingo” appears to be the Latin equivalent of the English word “feign.” In the 1729 English translation of the Principia, Andrew Motte translated {\it Hypotheses non fingo} as “I frame no hypothesis.” However, as Bernard Cohen stated, Newton never used the verb “frame” and besides it is not applied to hypotheses. For more discussion on this, see A Guide to Newton’s Principia, in Cohen and Whitman 1999 translation of Newton (1726), p. 274.}  For whatever is not deduced from the phenomena must be called a hypothesis; and hypotheses, whether metaphysical or physical, or based on occult qualities, or mechanical, have no place in experimental philosophy. In this philosophy particular propositions are inferred from the phenomena, and afterwards rendered general by induction.”
\smallskip

In Table 1, Newton's original statement is compared with the translations from Cohen and Withman (1999) and Châtelet (1759).

Because the General Scholium served as a recapitulation of all the arguments in the main text of the {\it Principia} to prove that celestial phenomena are not compatible with Descartes’s “hypothesis of vortices,” Cohen (1962) believed that Newton wrote this section to answer his critics, notably the Cartesians and other strict adherents of his mechanical philosophy. 

It is improbable that Euler was referring to Newton or to {\it Hypotheses non fingo} in particular. Rather, it is possible that he referred to other (unknown) philosophers who attempted to distance themselves from the Cartesian school to such extreme as to reject hypotheses altogether. In fact, Châtelet herself wrote that others had fallen into the opposite excess [of the Cartesians] who were disgusted with the suppositions and errors with which they found the books of Philosophy filled, and thus they had risen up against hypotheses, and had made them suspect and ridiculous, by calling them, {\it the poison of reason}, and {\it the plague of Philosophy}.\footnote{Châtelet (1742), §55.}

However, Châtelet concurred with Euler that hypotheses in physics are important. She believed that there must be a beginning in all research, and this beginning must almost always be a very imperfect attempt, and often without success. Émilie du Châtelet wrote a most eloquent statement about it:

\smallskip

{\it … celui-là seul qui serait en état d’assigner \& de démontrer les causes de tout ce que nous voyons, serait en droit de bannir entièrement les hypothèses de la Physique ; mais pour nous autres, qui ne semblons pas faits pour de telles connaissances, \& qui ne pouvons souvent arriver à la vérité qu’en nous traînant de vraisemblance en vraisemblance, il ne nous appartient pas de prononcer si hardiment contre les hypothèses.}\footnote{Ibid. §. 55.}

\smallskip

In 1759, ten years after her death, Clairaut published Châtelet’s translation of Newton’s {\it Principia}, appearing with the title {\it Principes mathématiques de la philosophie naturelle, par M. Newton (traduit du latin) par feu Madame la marquise du Châtelet}.\footnote{Châtelet 1759.}  She made no fuss over Newton’s {\it hypotheses non fingo}, and her translation into French is clear and succinct.\footnote{Ibid. Tome II, p. 179.}

\smallskip

Table 1. On Hypotheses by Newton and his translators.\\
\begin{center}
\centering 
\begin{tabular}{|c| c|} 
\hline 
Newton (1713) & {\it Rationem vero harum Gravitatis} \\ 
   & {\it  proprietatum ex Phœnomenis} \\ 
   & {\it nondum potui deducere, \&} \\ 
   & {\it Hypotheses non fingo.} \\ 
\hline
Cohen-Withman (1999)	 & I have not as yet been able to deduce \\ 
   & from phenomena the reason for these \\ 
   & properties of gravity, and \\ 
   & I do not feign hypotheses. \\ 
\hline
Châtelet (1759) & {\it Je n’ai pu encore parvenir à déduire des} \\ 
   & {\it  phénomènes la raison de ces propriétés} \\ 
   & {\it la gravité, \& } \\ 
   & {\it je n’imagine point d’hypothèses.} \\ 
\hline 
\end{tabular}
\end{center}

\subsection{Matière Subtile}

Châtelet did not address {\it matière subtile} (subtle matter), as Euler inferred in his undated note. In her discussion of phenomenal world, she argued that matter was to be described {\it in terms of extension, force, and inertia}. Extension combined with the passive force of inertia and the active moving force of {\it vis viva} is what we call matter. However, terms of her analysis were often inconsistent. First she stated that “the nature of a body consists of these three principles, ‘subsisting together’ and mutually independent.”\footnote{Châtelet 1740. §. 145, p. 159.}  Then she wrote that force is ‘different from matter’, but ‘inseparably attached to it’.\footnote{Ibid. §. 149, p. 164, 165.}  This is not what Euler was referring to. 

Euler specifically referred to Musschenbroek (not mentioned in the {\it Institutions}), criticizing the manner in which the experimental physicist reasoned on the phenomena of nature, and especially focusing on his explanation of {\it matière subtile}. Euler stated that “it is true that we cannot make the existence of such a matter seen by any experiment, but on the other hand absolutely deny the existence of any matter, as he does, of which we cannot see by the senses, it is much worse, than any hypothesis that we have made so far.”

Châtelet adhered to Leibniz’s concept of monads, an idealistic view that all substances are simple un-extended substances, the ultimate elements of the universe. On the other hand, Newton believed in “a most subtle Spirit which pervades and lies hid in all gross bodies; by the force and action of which Spirit the particles of bodies mutually attract one another at near distances” – to wit, an all-pervading gravitational ‘aether’.\footnote{Cf. Toni Vogel Carey 2012.}

Euler was an opponent of Leibniz’s theory of monads, firmly rejecting the teachings of Christian Wolff and his followers, and thus it is not surprising that he would find a delicate way to criticize Châtelet’s coverage of monads and adherence to Wolff’s teachings. Years later, Euler explained his views of Leibniz’s theory of “pre-established harmony” ({\it harmonie préétablie}), which deals with the relation between mind and matter, in several sections of his famous “Letters to a German Princess.” Pre-established harmony is a philosophical theory about causation under which every “substance” affects only itself, but all the substances (both bodies and minds) in the world nevertheless seem to causally interact with each other because they have been programmed by God in advance to “harmonize” with each other. 

Another perplexing part of Euler’s letter to Châtelet is his reference to magnetism and the experiments carried out by Musschenbroek, again a topic not found in her {\it Institutions}. Euler wrote:

{\it Ce principe porte même cet Auteur [Musschenbroek] si loin, qu’il ne doute pas attribuer les effets de l’aimant à un esprit, ou du moins à une substance incorporelle. Mais il me semble qu’on serait bien plus en droit d’exiger des expériences pour prouver l’existence de ces substances incorporelles plutôt que celle d’une matière subtile, qui est d’elle même si probable, que je n’en sçaurois douter.}\footnote{See Letter 2 in Appendix A.}

For Euler, physics was a quantitative theory based on sophisticated mathematics. He did work on magnetism and, being a mathematician, he was suspicious of alleged observations [by experimentalists] attempting to explain physical phenomena without analysis. 

Four years earlier, Daniel Bernoulli had written to Euler about Musschenbroek and his experiments with magnets. Daniel shared that Musschenbroek in Utrecht had made a new discovery that he was very proud of, namely that magnetic attraction decreased with the fourth power of distance. Euler responded to Bernoulli on 3 January 1739 (23 December 1738) with skepticism about Musschenbroek’s claim adding that Prof. G.W. Krafft (in Russia) had made experiments with a large magnet and that the magnetic attraction was far from reciprocal to the fourth power of the distances.\footnote{Euler 2016. Letters 35 and 36, pp. 284, 292.}

Hence, in 1742, Euler was well acquainted with Musschenbroek’s attempt to establish a ‘law of nature’ that governed magnetic phenomena by looking for a mathematical relation between the force of magnetism and its relation to other relevant quantities. In his {\it Dissertatio physica experimentalis de magnete} (1729), Musschenbroek stated that in the case of attraction one could show by experiments that the ‘cause of motion’ ({\it causa motus}) exists, but that we cannot know the modus operandi of this cause. Moreover, Musschenbroek attempted to establish a mathematical proportion that holds between the magnetic force and the minus-nth power of the distance on a par with the law of universal gravitation. Of course, he did not (could not) find a general magnetic law comparable to the law of universal attraction.

Perhaps Euler was attempting to make his views known about the importance of having a deep insight of mechanics since, as he told Goldbach, “without this understanding one commonly falls back upon mere illusions and contradictory hypotheses, which can neither consist with the true principles of Physics nor satisfy the phenomena.”\footnote{Euler 2015. Letter 78, pp. 297, 830. Euler to Goldbach, 25 April 1744.}

\subsection{On Forces and Principles of Mechanics}

Euler had strong ideas regarding the concept of force and the first law of mechanics, and in the letter to Châtelet he wrote: “I begin with the first principle of Mechanics that every body by itself remains in its state of rest or motion. To this property we may well give the name of force, when we do not say that every force is a tendency to change the state, as Mr. Wolf [sic] does. Every body is therefore provided … ”\footnote{See Letter 2 in Appendix A.} Unfortunately, the letter ends at this point. However, the incomplete sentence suggests that Euler intended to expand on the nature of forces. Would his explanation allude to {\it vis viva}, as Châtelet expected?

In his {\it Mechanica} (1736), Euler defined inertia as “{\it Vis inertiae est illa omnibus corporibus insita facultas vel in quiete permanendi, vel motum uniformiter in directum continuandi}.”\footnote{Euler 1736. Definitio 9, §74, Ch. 1.} [The force of inertia is that capacity inherent in all bodies either to remain at rest or to continue moving uniformly in a straight line.]  

In her {\it Institutions}, Châtelet introduced the law of inertia ({\it premier Loi}), and stated: “By the same principle of sufficient reason, a body in motion would never cease to move, if some cause did not stop its motion, by consuming its force; because matter also resists by its inertia to movement, when it is at rest, and at rest, when it is in motion.”\footnote{Châtelet 1742, §228.} Later, she wrote that “This effect is a necessary consequence of the first law of motion, based on the force of inertia of matter."\footnote{Ibid. §287.} 

Regarding the laws of motion ({\it loix générales du mouvement}), Châtelet stated that the active force and the passive force of the bodies, are modified in their collision, according to three main Laws:
\bigskip

PREMIERE Loi. {\it Un Corps persévère dans l’état où il se trouve, soit de repos, soit de mouvement, à moins que quelque cause ne le tire de son mouvement, ou de son repos}. [A body perseveres in the state in which it finds itself, either of rest or of motion, unless some cause draws it from its motion, or from its rest.]

SECONDE Loi. {\it Le changement qui arrive dans le mouvement d’un Corps, est toujours proportionel à la force motrice qui agit sur lui ; \& il ne peut arriver aucun changement dans la vitesse, \& la direction du Corps en mouvement que par une force extérieure ; car sans cela ce changement se seroit sans raison suffisante}. [The change which occurs in the motion of a body is always proportional to the motive force which acts upon it; \& there can be no change in the speed \& direction of the body in motion except by an external force; for without that this change would have been without sufficient reason.]

TROISIEME Loi. {\it La réaction est toujours égale à l’action ; car un Corps ne pourroit agir sur un autre Corps, si cet autre Corps ne lui resistoit : ainsi, l’action \& la réaction sont toujours égales \& opposées}. [Reaction is always equal to action; for a body could not act on another body if this other body did not resist it: thus, action and reaction are always equal and opposite.]
\bigskip

Euler continued expanding his research to establish more clearly the nature of forces. In 1752, he published {\it Recherches sur l’origin des forces}\footnote{Euler 1752.} where he redefined inertia and argued that impenetrability was one of the basic characteristics of matter. In 1982, Stephen Gaukroger carried out a study on “The Metaphysics of Impenetrability,” focusing mainly on Euler’s conception of force.\footnote{Gaukroger 1982.}  Gaukroger concluded that “The only justification for the principle of inertia that we are given in Euler’s Theoria, and indeed the only justification he gives is in terms of the principle of sufficient reason is this: a body will not change its state without sufficient reason, where the sufficient reason is specified in terms of external forces."

\section{Opportunity Lost}

Euler was an expert in mechanics, and his mathematical prowess was firmly established when Châtelet needed guidance. Euler (Fig. 3) was the first to apply the full analytic power of the calculus to the study of motion. Although Newton’s {\it Principia} was fundamental to the understanding of mechanics, it lacked in analytical sophistication, i.e. the mathematics required to explain the physics of motion was rather obscure, as Newton had preferred to use geometrical arguments rather than his own invented calculus. With his two-volume {\it Mechanica}, Euler described analytically the laws governing movement and presented mechanics in the form of mathematical analysis for the first time.\footnote{Musielak 2022.} Later, he published another mechanics text in which he introduced the equations for rigid body rotations.

\begin{figure}
\begin{center}
\includegraphics{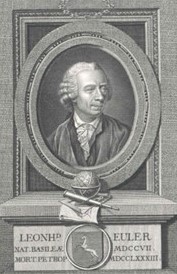}
\end{center}
\caption{Undated Portrait of Leonhard Euler (1707-1783)}
\label{figure}
\end{figure}

Euler arrived in Berlin in the summer of 1741. The incomplete message he wrote to Châtelet can be dated to May 1742 or later, after she published the second edition of {\it Institutions}. It is unknown whether Euler ever finished and sent this letter. There is a missive in the Russian archives written by Châtelet on 30 May 1744 where “she dwells on her dispute with de Mairan and polemizes with Euler on the principle of living forces.”\footnote{Smirnov 1963, p. 279.} This suggests that either she did not receive his message in 1742, or the ideas he shared did not match her expectations. Of course, by 1744 the dispute with Mairan had run its course and she no longer needed an advocate.

There is no record that Euler sought Châtelet or that she visited Maupertuis in Berlin. Had they met in person, it is doubtful that Émilie would have liked Euler. For one thing, Euler was a pious, simple family man, too unsophisticated and unpolished for her taste. Émilie du Châtelet preferred men like Voltaire – literary, witty, sophisticated – or elegant scholars like Maupertuis, Clairaut, and the young Bernoulli, gentlemen who were charmed by her own wittiness. Euler and Madame du Châtelet belonged to two completely different worlds. Besides, Euler’s physical appearance may have turned her off even more.\footnote{When Axel Freiherr von Mardefeld wrote to the Prussian king on 17 June 1741, he referred to Euler as “le plus grand Algébriste de l’Europe”, telling him that the Russian court regrets the loss – and warned at the same time that Euler’s physical appearance did not give a favourable impression [referring to his blind right eye, perhaps].} His blind right eye was shrinking since he lost vision to a grave illness several years earlier.

In any case, had their relationship developed, Euler could have helped Châtelet in her exposition of mechanics when she worked on the translation of Newton’s {\it Principia}. 

\section{Euler's Derivation of Vis Viva, Momentum, and Effectiveness (Energy) of Motion}

Euler did not publish his analysis of {\it vis viva}, and the undated letter only reveals his frustration that “that this matter had already been the subject of such a bitter dispute for so long." However, in a treatise discovered in 1844, Euler delves into the subject and offers a clear and important explanation of the difference between momentum and {\it vis viva}, and he defined a new concept that led him to an early representation of kinetic energy. 

\subsection{Euler's (Unpublished) Introduction to Natural Science (Physics)}

Euler carried out an analysis to distinguish the two debated concepts in his “Introduction to Natural Science,"\footnote{Euler 1862.} a manuscript that could have been the first comprehensive treatise of physics based on mathematics, if he had published it. Written in German with the title {\it Anleitung zur Natur-Lehre, worin die Gründe zu Erklärung aller in der Natur sich ereignenden Begebenheiten und Veränderungen festgesetzet werden}, the twenty-one chapter book addresses important concepts in natural science [physics], which he defined as “the science that aims to explain the causes of the changes that occur on material bodies." Euler deals with the most misunderstood ideas of his time, starting from the concepts of mass, density, and weight, the general properties of bodies, followed by the effects of forces on the velocity of bodies, determination of motion of a body driven by different forces, and the laws of statics and dynamics of liquid matter.

While covering the effects of forces on the velocity of bodies, Euler stated that when a moving body of mass $M$ is propelled by a force $p$, the increase in its velocity produced in a certain time is as the force multiplied by the time [increment] and divided by the mass of the body. In equation form, he wrote this as $dv = \frac{npdt}{M}$, where $n$ denotes a constant that depends on the units of measurement.\footnote{Ibid. §56, Ch. VII.}  I rewrite this expression in the more familiar version of the second law, in which a force $F$ produces an acceleration or a change in momentum in a given time: 

\begin{equation}
M\frac{dv}{dt} = F
\end{equation}

where $dv$ is the increment of velocity, and $F$ is the force applied to body of constant mass $M$. This is the expression for the momentum law (Newton's second law of mechanics). It is important to note that Euler had already introduced this momentum expression in {\it Mechanica} (1736).

In the same chapter VII, Euler explains the difference between $Mv$ (momentum) and $Mv^2$ ({\it vis viva}). He sets the stage as follows: Let $M$ be the mass of a body, $v$ its speed, and $p$ the backward force acting on it, gradually reducing its speed, $t$ the time in which the body is brought to complete rest, and $s$ the path that it traverses in that time. Then, because a backwards acting force reduces the speed by the same amount as it would increase it, if it acted on the body in a forward direction, the following two equations would hold: $Mv = npt$ and $Mv^2 = 2nps$. I rewrite these expressions in modern form:

\begin{equation}
M\frac{dv}{dt} = F; \hskip0.2in  \frac{1}{2}Mv^2 =sF 
\end{equation}

Euler explains that, if two different bodies in motion are to be brought to rest at the same time, the forces required must vary as $Mv$, i.e. as the magnitudes of motion of the bodies. But if the same bodies were to be brought to rest not in the same time, but whilst traversing the same path, the required forces would have to vary as $Mv^2$, that is, as the masses multiplied with the square of the speed, or as their so-called living forces. According to Euler, this was the cause of the dispute which arose from the belief of some people that the force of a moving body must be determined from the product of its mass with its speed [momentum], whilst others maintain that the force must be determined from the product of the mass and the square of the speed [{\it vis viva}]. Euler believed that the misunderstanding arose from the need that one wants to associate an actual force with a moving body, although neither $Mv$ nor $Mv^2$ can be compared with a force, since in one case the force must be connected with time, and in the other with the path.

In {\it Anleitung zur Natur-Lehre}, Euler also derived a new concept he called “Effectiveness" [{\it Wirksamkeit}] of motion. First he declared that whatever the nature of the forces acting on the body are, they can always be decomposed into three, which push the body straight away from each of the assumed three planes, and the action of each is not affected by the remaining two other components; he denoted the components of the force along the three Cartesian axes as $P$, $Q$, and $R$, respectively, which for clarity here we write in the familiar form $F_x$, $F_y$, and $F_z$. Then Euler obtained the integral quantity by the effectiveness of motion when a force acts on body of mass $M$, which he found by multiplying the force by the differential of its distance from the plane from which it pushes the body away. He obtained integral equations for a quantity that could now be called “energy of motion." With the three components of velocity denoted $u$, $v$, and $w$, and the components of the force expressed as $F_x$, $F_y$, and $F_z$  along the three axes $x$, $y$, $z$, Euler found the "effectiveness of motion" by integration:

\begin{equation}
\frac{1}{2}Mu^2 = \int F_{x}dx; \hskip0.11in \frac{1}{2}Mv^2 = \int F_{y}dy; \hskip0.11in \frac{1}{2}Mw^2 = \int F_{z}dz
\end{equation}

These expressions appear as a statement of what now we call 'kinetic energy' of a body of mass $M$. Euler remarked: “This concept is of great importance, because the sum of the effectiveness of all forces $\int F_{x}dx + \int F_{y}dy + \int F_{z}dz$ always has the same magnitude, even if we had assumed three different planes; and if the three forces have arisen from the decomposition of a single force, then the sum of their effectiveness is equal to the effectiveness of the single force, from which they were derived."\footnote{Ibid. §75, Ch. IX.} 

Euler ended the explanation of his discovery by stating that “what makes this concept of effectiveness in itself most remarkable is the fact that the whole theory of equilibrium is based on it." He added that it can be shown that equilibrium can not occur, if the sum of the effectivenesses is not a minimum, or occasionally a maximum. He remarked: “This wonderful principle was also first conceived by the world-famous President von Maupertuis, and is closely related to the other general law of economy [{\it Verbindung}]. From this we see at least that the effectiveness has an essential influence on all motions that can be brought about by forces, and that it certainly deserves to be given a special name."\footnote{Ibid. §75, Ch. IX.}

Although he gives credit to Maupertuis, in 1743 Euler had already demonstrated that for motion of a point mass under the influence of a central force the formula $\int Mv ds$, where $M$ denotes the mass, $v$ the velocity and $ds$ the element of space covered, is always a minimum.\footnote{Euler 1744, Additamentum II.}  In other words, the corresponding trajectory minimizes the action integral $\int Mv ds$. 

I believe that Euler composed his “Introduction to Natural Science" shortly after his exchange with Châtelet, probably around 1745. This assumption is supported by the fact that Euler aluded to Maupertuis and the principle of least action, a topic that he had already addressed in 1743. After reading Châtelet's excellent {\it Institutions}, Euler may have felt the need to express his own description of natural science (physics), using analysis to explain the physical concepts that were still unclear and thus contested. Of course, this assumption requires further investigation. It is not clear why he did not publish this work.

\subsection{End of Debate}

What issues were resolved in the Châtelet-Mairan dispute? None. Neither contender solved the issues raised. Because the concepts of momentum, work, and energy were not well understood, the idea of {\it vis viva} was somehow clouded by individual’s understanding of mechanical principles. However, Émilie du Châtelet developed Leibniz’s concept and, combining it with s’Gravesande’s experimental observations,\footnote{{\it Physices elementa mathematica, experimentis confirmata. Sive introductio ad philosophiam Newtonianam} / [Willem Jacob ‘s Gravesande], 1720-1721.} showed that {\it vis viva} was dependent on the square of the velocities. She attempted to integrate aspects of Cartesianism and Newtonianism with Leibnizian ideas. Her Leibnizian convictions in dynamics and her lack of mathematical training prevented her from recognizing the validity of both measures of ‘force’, $mv$ and ‘energy’ $mv^2$. Now we say that {\it vis viva} is an early representation of kinetic energy. But it was not until early nineteenth century when {\it vis viva} was viewed as a manifestation of energy.

The end of the {\it vis viva} disputes can be attributed to Euler. As Marius Stan has stated eloquently: “In 1754, Euler was single-handedly extending Newton’s Lex Secunda to fluids, thus making it the fundamental principle of classical mechanics, with the Second Law having outlived and displaced Conservation of Vis Viva.”\footnote{Stan 2017, p. 289.}

Today, we can define momentum of a body as force $F$ acting through a finite time $t$: since velocity is $v = at$, we can write for a body of mass $m$ the momentum is $mv = mat = Ft$. Moreover, momentum is the quantity of motion of a moving body, namely a vector. In a collision, momentum is conserved.

The kinetic energy of a body can be defined as force acting over a space $s$, and since $v^2 = 2as$ for a body of constant mass $m$ we write $mv^2=2mas$, which is equivalent to $\frac{1}{2} mv^2 = Fs$. Now we say that kinetic energy (KE) is the energy which a body possesses by virtue of being in motion, and we know it is a scalar quantity. In a collision, KE is not conserved.

Confusion over these two terms was apparent in the various arguments of the debaters in the eighteenth century. Eventually, it became clear that both concepts are equally valid in the description of motion. It would take many years and new mathematical developments to resolve these issues. In fact, the study of motion carried out by Euler, and later by Lagrange, and others after focused on specific problems and on principles from which their mathematical solutions could be derived. And now these physical concepts are clearly defined.

\bigskip

{\bf Acknowledgement}\\

 {\it Nihil turpius est physico, quam fieri sine causa quicquam dicere.}\\

\bigskip
\bigskip

{\bf References}\\

Brown, H. (1975). From London to Lapland: Maupertuis, Johann Bernoulli I, and La Terre Applatie, 1728-1738. In Literature and History in the Age of Ideas, Charles G. S. Williams, Ed. Ohio State University Press (1975).

Châtelet, É. [anon.] (1738-1740). Institutions de physique. Paris: Prault. Manuscript Bibliotheque nationale de France Paris, Fonds français 12265. 
 
Châtelet, G.-É. marquise du (1742). Institutions physiques de madame la marquise du Chastellet adressées à M. son fils. Nouvelle édition, corrigée et augmentée, considérablement par l’auteur, Amsterdam, Aux dépens de la Compagnie, 1742.

Châtelet, É. (1818). Lettres inédites de Mme la Mise Du Châtelet, et supplément à la correspondance de Voltaire avec le roi de Prusse et avec différentes personnes célèbres on y a joint quelques lettres de cet écrivain, qui n'ont point été recueillies dans les Oeuvres complètes avec des notes historiques et littéraires, par Sérieys et Eckard.

Châtelet, É. (2015). Newton, Isaac : Principes mathématiques de la philosophie naturelle. La traduction francaise des Philosophiae naturalis principia mathematica, 2 vols. Édition critique du manuscript par Michel Toulmonde, Ferney-Voltaire: Centre international d’étude du XVIIIe siècle. First published 1759.

Cohen, I. B. (1962). The First English Version of Newton’s Hypotheses non fingo. Isis, Sep., 1962, Vol. 53, No. 3 (Sep., 1962), pp. 379-388.

Davidson, I. (2012). Voltaire: A Life. Pegasus. Reprint edition, March 12, 2012.

Descartes, R. (1644). Principia philosophiae. Oeuvres de Descartes, ed. C. Adam and P. Tannery, 13 vols. (Paris: Cerf, 1897-1913), Vol. VIII, p. 61.

Euler, L. (1736). Mechanica sive motus scientia analytice exposita. (Mechanics or Exposition of Analytic Science of Motion). Typographia Academiae Scientiarum: Petropoli, 1736 – Opera Omnia II/2 [E. 15, E. 16].

Euler, L. (1738). Dissertatio de igne, in qua eius natura et proprietates explicantur (Dissertation on fire). Recueil des pièces qui ont remporté les prix de l’académie royale des sciences 4, 1752, pp. 3-19. Opera Omnia: Series 3, Volume 10 [E. 34] 1738. 

Euler, L. (1744). Methodus inveniendi curvas h’neas maximi minimive proprietate gaudentes sive solution problematis isoperimetrici latissimo sensu accepti. Lausanne, Genf: M.-M. Bousquet. Reprinted in Euler, L. Opera Omnia, I 24. (E. 65). According to Eneström, Euler completed this manuscript by April 1743.

Euler, L. (1752). Recherches sur l’origine des forces (Research on the origin of forces). Mémoires de l’académie des sciences de Berlin, Volume 6, 1752, pp. 419-447.

Euler (2015). Leonhardi Euleri Opera Omnia. Series Quarta A : Commercium Epistolicum. Volumen quartum: Correspondance de Leonhard Euler with Christian Goldbach. Springer Basilea MMXV. Springer Basel 2015.

Euler (2016). Leonhardi Euleri Opera Omnia. Series Quarta A : Commercium Epistolicum. Volumen quartum: Correspondance de Leonhard Euler with Daniel, Johann II, Johann II Bernoulli.  Birkhäuser Basilea MMXVI. Birkhäuser 2016.

Gauker, S. (1982). The Metaphysics of Impenetrability: Euler’s Conception of Force. The British Journal for the History of Science, Vol. 15, No. 2 (Jul., 1982), pp. 132-154.

Hagengruber, R. (2011). Émilie du Châtelet between Leibniz and Newton: the Transformation of Methaphysics. in Émilie du Chatelet between Leibniz and Newton, edited by Ruth Hagengruber. Springer, 2011.

Hamel, F. (1910). An Eighteenth Century Marquise: a study of Émilie du Châtelet and her times. London: Stanley Paul and Co.

Leibniz, G. W. (1686). Brevis demonstratio erroris memorabilis Cartesii et aliorum circa legem naturalem, secundum quam volunt a Deo eandem semper quantitatem motus conservari; qua et in re mechanica abutuntur, Acta Eruditorum, 1686, pp. 161-163. A translation in Gottfried Wilhelm Leibniz, Philosophical Papers and Letters, trans. Leroy E. Loemker, 2 vols. Chicago: Univ. of Chicago Press, 1956, Vol. I, pp. 455-463.

Musielak, D. (2014). The Marquise du Châtelet: A Controversial Woman of Science, Arxiv.org, ID: 1010553.

Musielak, D. (2022). Leonhard Euler and the Foundations of Celestial Mechanics.  Springer History of Physics book series. ISBN: 978-1-119-64060-8. 

Newton, I. (1726). Philosophiae Naturalis Principia Mathematica. English translation by I. Bernard Cohen and Anne Whitman, University of California Press, 1999. ISBN 0-520-08817-4.

Reichenberger, A. (2018). Émilie Du Châtelet’s interpretation of the laws of motion in the light of 18th century mechanics. Studies in History and Philosophy of Science 69 (2018), 1-11.

Smirnov (1963). T.N. Klado, Yu.K. Kopelevich, T.A. Lukina (compilers); V.I. Smirnov (ed.), [Leonhard Euler: Letters to scholars], Akademiya Nauk, 1963.

Smith, G.E. (2006). The vis viva dispute: a controversy at the dawn of dynamics, Physics Today (October), 31-36 (2006).

Stan, M. (2017). Newton’s Concepts of Force among the Leibnitzians. In Reading Newton in Early Modern Europe. Edited by Elizabethanne Boran and Mordechai Feingold, Brill Academic Pub (June 15, 2017).

Terral, M. (2004). Vis Viva Revisited. Hist. Sci. XLII (2004).\\

\break



{\bf Appendix A: Letters from Euler to Châtelet}

\bigskip

The following letters are preserved at the Leningrad Archive of the Soviet Academy of Sciences in a file containing copies of Euler's epistle to Clairaut for the same period. These letters were published in a volume edited by Smirnov (1963) in the original French language and translated into Russian.

\bigskip
\bigskip

Letter 1 – Euler to Châtelet (19 February 1740)

\bigskip

Madame,

Quoique je n’aye pas encore reçu la Pièce{\footnote {{\it Institutions de Physique} (1740). See my explanation in Section 3.}}  que Vous avez eu la bonté de remettre à Mr. de Maupertuis pour moi, je ne puis différer de prendre la liberté de Vous remercier d’une distinction aussi avantageuse. Il m’est bien glorieux, Madame, de me voir en lice{\footnote{French Academy prize for 1738 on the nature and propagation of heat. See my description in Section 2.2.}}  avec une personne qui fait un des plus rares ornements de son Sexe, par le Lustre que Vous avez bien voulu répandre sur les sciences les plus relevées, en y portant la Sublimité de Votre génie. Cela seul, je le sens bien, est capable de me donner un Relief que je n’oserais espérer de mes faibles lumières.{\footnote{I translate {\it mes faibles lumières} as “my modest knowledge.” Euler, who was thirty-three years old, was clearly awestruck by Châtelet’s aristocratic lineage and intellectual brilliance and therefore he used a rather self-deprecating term of humility. }}  Si j’avais cependant lieu de me féliciter de quelque chose à cet égard, ce serait uniquement à cause de l’honneur qu’elles viennent de me procurer et pour être en état, Madame, de Vous admirer avec connaissance de cause. Il me tarde beaucoup de recevoir Votre ouvrage{\footnote{In February 1740, Euler had not read her {\it Institutions de Physique} (first edition) and perhaps he never did, until he received the second edition in May 1742. See Section 4.5.}}  pour Vous rendre ce tribut que tous ceux qui aiment les sciences doivent se faire un honneur et un plaisir de Vous rendre. 

Agréez, Madame, la sincérité du profond respect avec lequel j’ai l’honneur d’être, Madame,
Votre très humble et très obéissant serviteur

Leonhard Euler\\
\bigskip

Letter 2 – Euler to Châtelet (undated) – evidence suggests it was written after May 1742.

\bigskip

Madame,

L’honneur dont Vous me daignez, en me communiquant Vos excellents ouvrages, me met dans une si grande confusion, que je ne sais pas Vous exprimer ma reconnaissance. En lisant vos Institutions Physiques, j’ai également admiré la clarté, avec laquelle Vous traitez cette science, que la facilitée, avec laquelle Vous expliquez les choses les plus difficiles sur le mouvement, qui sont même assez embarrassantes, quand il est permis de se servir du calcul. Mais surtout le Chapitre sur les hypothèses{\footnote{Hypotheses of Physics. See Section 5.1.}} m’a fait le plus grand plaisir, voyant, que Vous combattez, Madame, si fortement et si solidement quelques Philosophes Anglais, qui ont voulu bannir tout à fait les hypothèses de la Physique  qui sont pourtant à mon avis le seul moyen de parvenir à une connaissance certaine des causes physiques. J’ai été souvent en peine, lorsque le Discours rouloir sur cette matière avec des Anglois, de trouver des raisons convaincantes, pour bien faire voir l’utilité des hypothèses, mais je n’ai jamais pu développer mes idées sur ce sujet d’une manière si claire, que Vous. J’estime Mr Musschenbroek comme un très grand physicien et j’ai beaucoup profité de ses expériences, mais sa manière de raisonner me fait un si grand dégoût, qu’à peine je me puis résoudre à lire ses raisonnements, qu’il fait sur les phénomènes de la nature.

Ce grand homme se met entièrement en colère,{\footnote{By writing {\it Ce grand homme se met entièrement en colère} (“This great man becomes entirely unreasonable when he speaks of subtle matter”) Euler may be expressing his opposition to the idea of monads, as expounded by Wolff.}}  quand il parle de quelque matière subtile{\footnote{Subtle matter. See my discussion in Section 5.2.}}  que d’autres emploient pour expliquer plusieurs phénomènes : il est vrai qu’on ne peut pas faire voir l’existence d’une telle matière par aucune expérience, mais d’un autre coté nier absolument l’existence de tout matière, comme il fait, dont on ne se peut apercevoir par les sens, c’est bien pire, qu’aucune hypothèse, qu’on a fait jusqu’à présent. Ce principe porte même cet Auteur{\footnote{Musschenbroek and his studies of magnetism. See Section 5.2.}}  si loin, qu’il ne doute pas attribuer les effets de l’aimant à un esprit, ou du moins à une substance incorporelle.  Mais il me semble qu’on serait bien plus en droit d’exiger des expériences pour prouver l’existence de ces substances incorporelles{\footnote{Incorporeal substances. Section 5.2.}} plutôt que celle d’une matière subtile, qui est d’elle même si probable, que je n’en sçaurois douter. Mais j’espère qu’une bonne partie de ces gens changeront bientôt leur sentiment après avoir lu Votre admirable dissertation sur les hypothèses : et je ne doute nullement, que Mr. Demairan ne soit entièrement convaincu par les solides raisons que Vous avez opposées à ses idées si mal fondées sur la force des corps.{\footnote{Dispute with De Mairan, as described in Section 4.3.}} 

Au reste je plains fort, que cette matière ait été déjà si longtemps le sujet d’une dispute si forte, par laquelle la mathématique a perdu beaucoup de sa réputation. Et il me semble, que si l’on vouloir examiner cette matière depuis sa source véritable, on n’y trouverait la moindre difficulté. Agréez donc Madame, que je Vous présente mes pensées là-dessus. Je commence par le premier principe de la Mécanique que tout corps par lui-même demeure dans son état ou de repos ou de mouvement. A cette propriété on peut bien donner le nom de force,{\footnote{Concept of Force. See Section 5.3.}}  quand on ne dit pas que toute force est une tendance de changer l’état, comme fait Mr. Wolf. Tout corps est donc pourvu ...

\bigskip

$\hskip0.5in$                              {Dora Musielak; University of Texas at Arlington, January 2023}

\end{document}